\pdfoutput=1
\PassOptionsToPackage{prologue,dvipsnames}{xcolor}
\documentclass[11pt]{article}

\usepackage[]{acl}
\usepackage[dvipsnames]{xcolor}
\usepackage{times}
\usepackage{latexsym}
\usepackage{multirow}
\usepackage{graphicx}
\usepackage{amsmath,amsthm}
\usepackage{comment}
\usepackage{caption}
\usepackage{subcaption}
\usepackage{booktabs}
\usepackage{array}
\usepackage{tcolorbox}
\usepackage{color,soul}
\usepackage{enumitem}
\usepackage{longtable}
\usepackage{setspace}
\usepackage{multicol}
\usepackage[T1]{fontenc}

\usepackage[utf8]{inputenc}

\usepackage{microtype}

\usepackage{inconsolata}

\title{The Whole is Better than the Sum: Using Aggregated Demonstrations in In-Context Learning for Sequential Recommendation}
\author{
~~Lei Wang
~~Ee-Peng Lim\thanks{~~Corresponding author.} \\
Singapore Management University\\
{\tt \{lei.wang.2019, eplim\}@smu.edu.sg} \\
}

\begin{document}
\maketitle
\begin{abstract}
Large language models (LLMs) have shown excellent performance on various NLP tasks. To use LLMs as strong sequential recommenders, we explore the in-context learning approach to sequential recommendation.  We investigate the effects of instruction format, task consistency, demonstration selection, and number of demonstrations. As increasing the number of demonstrations in ICL does not improve accuracy despite using a long prompt, we propose a novel method called LLMSRec-Syn that incorporates multiple demonstration users into one aggregated demonstration. Our experiments on three recommendation datasets show that LLMSRec-Syn outperforms state-of-the-art LLM-based sequential recommendation methods. In some cases, LLMSRec-Syn can perform on par with or even better than supervised learning methods. Our code is publicly available at \url{https://github.com/demoleiwang/LLMSRec_Syn}.
\end{abstract}

\section{Introduction}
\label{sec:intro}

{\bf Motivation.}
Large language models (LLMs) are known to perform well as a zero-shot solution for many natural language processing tasks~\cite{brown2020language, chowdhery2022palm, openai-chatgpt-2022, qin2023chatgpt}. 
Recently, there are some works that focus on using LLMs to perform recommendation with promising accuracies~\cite{hou2023large, wang2023zero, liu2023chatgpt, bao2023tallrec, gao2023chat} and to provide explanations~\cite{yang2023large, wang2023llm4vis}.  Most of these works developed LLM prompts for zero-shot sequential recommendation.

\begin{figure}[t]
    \centering
    \includegraphics[width=0.48\textwidth]{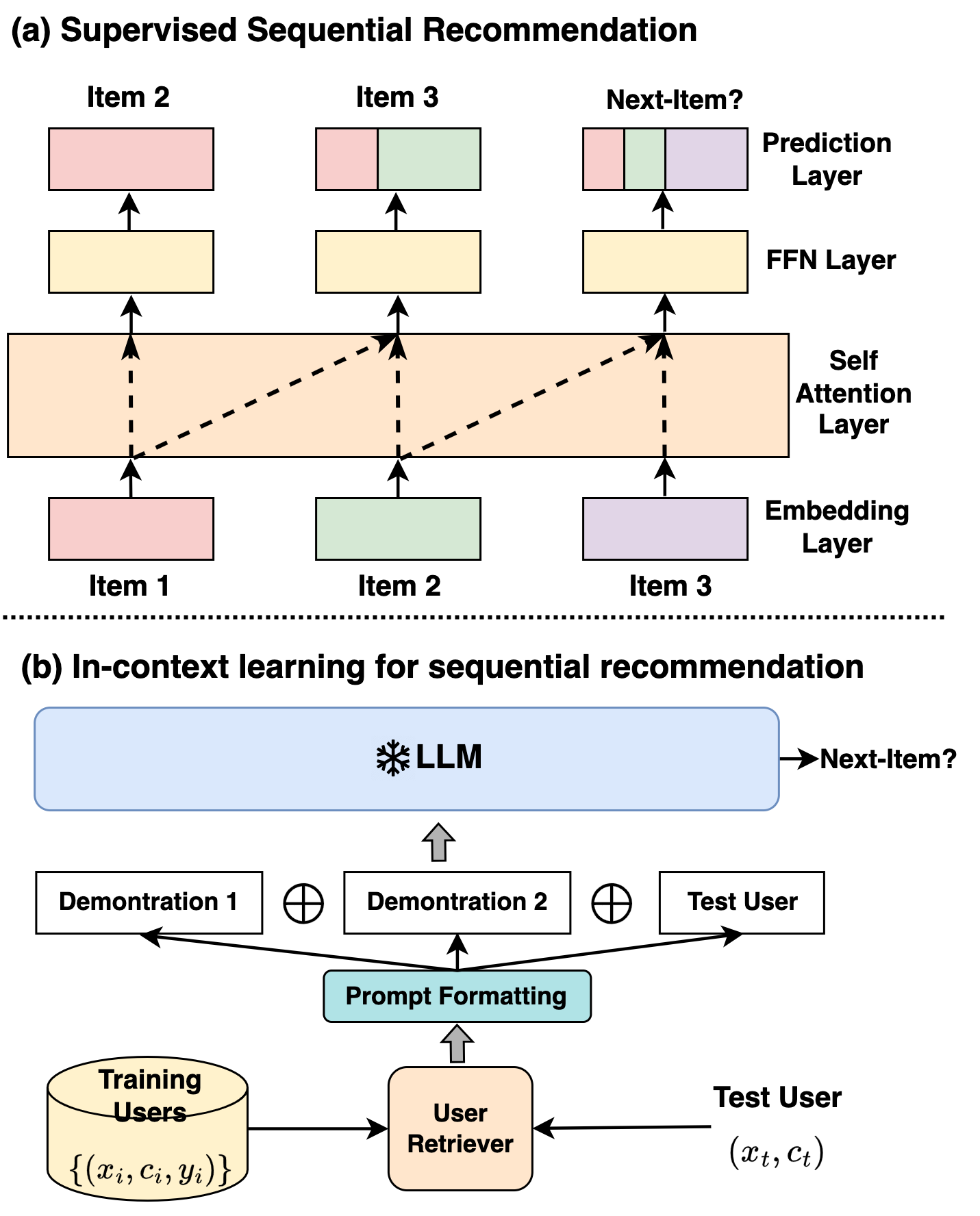}
    \caption{The illustrative comparison of (a) supervised sequential recommendation method and (b) in-context learning based sequential recommendation method.}
    \label{fig:method_comparison}
\end{figure}  

To investigate whether LLM can serve as a strong zero-shot sequential recommender, ~\citet{hou2023large} devised a prompt that is filled with historical items in chronological order, candidate items, and instruction to rank the candidate items. ~\citet{wang2023zero} proposed a three-step prompting method, where LLMs first summarizes the user preference based on the user's past interacted items. It then identifies representative items from the past interacted items that capture the user preference, and finally recommends items among the candidate items which are aligned with the representative items. Among the very few one-shot sequential recommendation works, \citet{liu2023chatgpt} and ~\citet{hou2023large} explored in-context learning using the test user's  second last item as the ground truth next-item and all earlier interacted items as input to create self-demonstrations.
Nevertheless, previous experiments have shown that in-context learning (ICL) based sequential recommendation methods perform poorly compared with the supervised learning-based methods (e.g., SASRec) due to the complex recommendation task definition~\cite{liu2023chatgpt, hou2023large, wang2023zero}. The illustrative comparison of these two methods is shown in Figure~\ref{fig:method_comparison}.

To develop an effective in-context learning approach for LLMs to perform sequential recommendation, we first define the sequential recommendation problem as follows.

{\bf Problem definition.}
We denote each input user instance $u_i$ to be a $(x_i,c_i,y_i)$ tuple where $x_i$ denotes the sequence of past interacted items (excluding $y_i$) by $u_i$, $c_i$ denotes the candidate items to be recommended ($|c_i|=M$), and $y_i$ denotes the ground truth next-item which is also the last item interacted by $u_i$.  Note that $y_i$ appears in $c_i$ ($y_i \in c_i$). 
A LLM-based sequential recommendation method is required to assign a rank $rank(d) \in [1,M]$ to each item $d$ in $c_i$.  
Our objective is to ensure that the method ranks $y_i$, i.e., $rank(y_i)$, as high as possible for all. 

The above definition includes $c_i$ as input as it is usually infeasible for LLMs to take all items as input due to limited prompt length.  Moreover, having $c_i$ does not introduce bias in the evaluation.  
The above definition is also adopted in ~\citet{hou2023large}. 
We also assume a dataset of users' interacted item sequences from which we can construct demonstrations for ICL, and a LLM which is too large for pretraining or finetuning.  





{\bf Overview of our study.} Past works has shown that the effectiveness of ICL in adapting LLMs to new tasks is significantly influenced by instruction wording~\cite{madaan2022text, yang2023large}, label design~\cite{yoo2022ground, wei2023larger}, selection of demonstrations~\cite{liu2021makes, shi2022xricl, zhang2023makes}, and number of demonstrations~\citet{chen2023many, zhao2023dynamic}.
Our study thus begins by systematically investigating how the instruction format, task consistency (between test and demonstration), demonstration selection, and the number of demonstrations affect ICL-based sequential recommendation.
Through our preliminary experiments, we obtain four findings including the one that observes degradation of recommendation accuracy when the number of demonstrations increases.  As each demonstration takes up significant length, it is also easy for multiple demonstrations to exceed the prompt limit of LLMs.  Moreover, as LLMs are known to miss out relevant information in a long input prompt~\cite{liu2023lost}, we thus embark on a follow-up study on designing a more efficient ICL scheme based on {\it aggregated demonstration}.

\begin{figure}[t]
    \centering
    \includegraphics[width=0.48\textwidth]{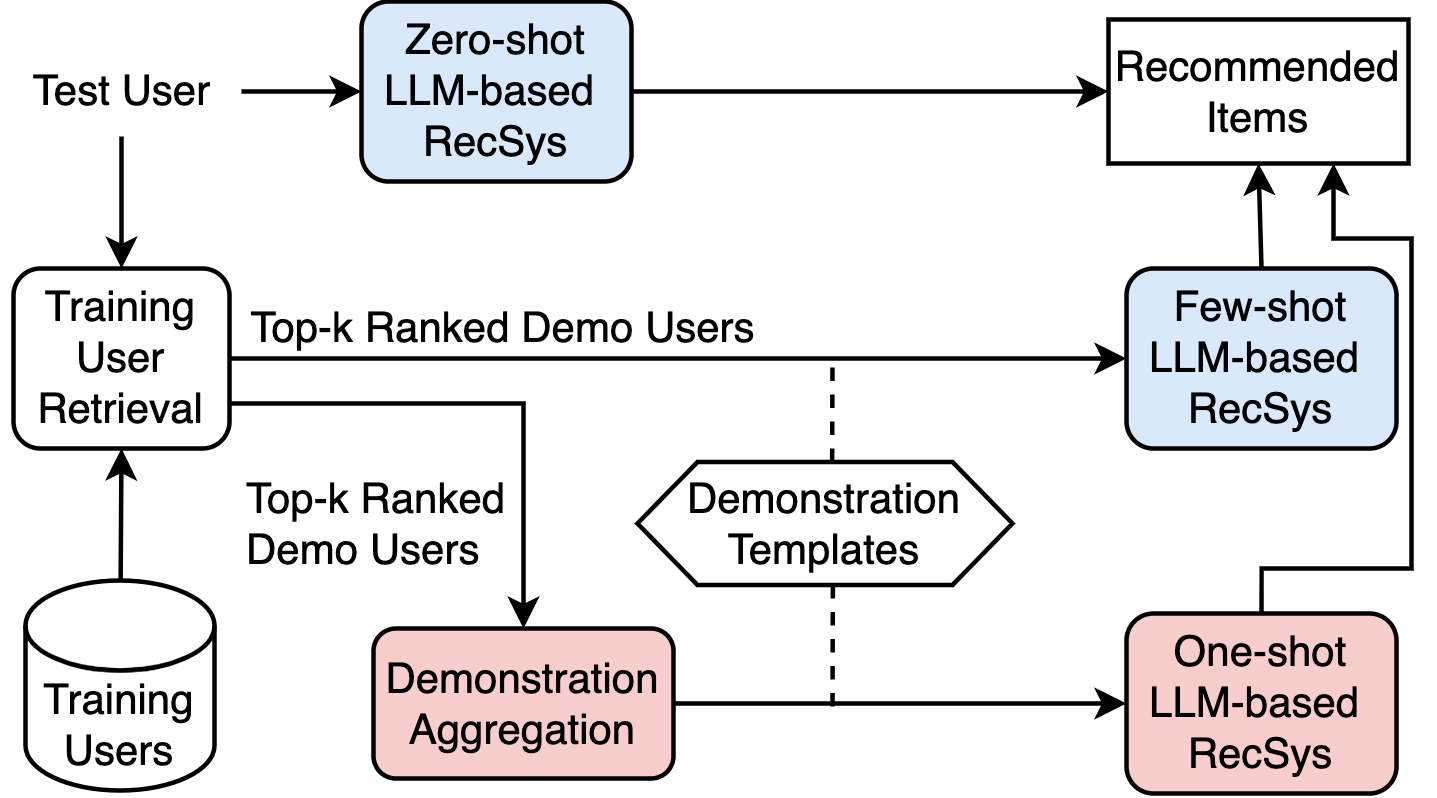}
    \caption{The overall framework of zero-shot, few-shot, and aggregated one-shot LLM-based sequential recommender systems.}
    \label{ch5_fig:framework}
\end{figure}

Figure~\ref{ch5_fig:framework} shows a comparison of the frameworks for zero-shot, few-shot, and aggregated one-shot
LLM-based sequential recommender systems. The key idea in aggregated demonstration is to combine multiple training users into one demonstration. This reduces the repetition of instruction text in the ICL prompt. It also seeks to summarize multiple training users relevant to the test instance in a compact manner.
We also develop a novel ICL method using aggregated demonstration for sequential recommendation known as \textbf{LLMSRec-Syn}. The length of LLMSRec-Syn prompt increases only gradually with number of demonstration users, LLMSRec-Syn can cope with more relevant information from the demonstration users within a concise input context. We finally show LLMSRec-Syn outperforms other zero-shot and one-shot ICL methods in an extensive set of experiments.

{\bf Contribution.} Our contributions can be summarized as follows: (1) We systematically explore the ICL approach to sequential recommendation by empirically investigating the effect of instruction format, task consistency, demonstration selection, and number of demonstrations; (2) We propose a new in-context learning method for sequential recommendation called LLMSRec-Syn which leverages on a novel concept of  aggregated demonstration; (3) We experiment on three popular recommendation datasets and show that  LLMSRec-Syn outperforms previous LLM-based sequential recommendation methods. 

\section{Related Work}

\textbf{In-Context Learning.}
Several works show that LLMs can effectively adapt to different NLP and multimodal tasks, including machine translation~\cite{agrawal2022context},
visual question answering~\cite{yang2022empirical}, and foreground segmentation~\cite{zhang2023makes}. This adaptation is achieved by learning from a few task-relevant demonstrations, commonly known as in-context learning (ICL)~\cite{brown2020language}.
Despite the above successes, ICL's performance is still significantly affected by the wording of instructions~\cite{madaan2022text, yang2023large}, label design~\cite{yoo2022ground, wei2023larger}, demonstration selection~\cite{liu2021makes, shi2022xricl, zhang2023makes}, and number of demonstrations~\citet{chen2023many, zhao2023dynamic}.   ICL is much less studied in LLM-based sequential recommendation. As sequential recommendation is distinct from the pretraining tasks of LLMs and also different from the above-mentioned tasks, new designs of demonstration(s) and ICL prompt is necessary.

\textbf{LLMs for Sequential Recommendation.}
Early sequential recommendation works adopt techniques such as Markov Chains~\cite{rendle2010factorizing, he2016fusing} and neural networks (e.g., RNN~\cite{hidasi2015session}, CNN~\cite{tang2018personalized}, Self-Attention~\cite{kang2018self}, and GNN~\cite{chang2021sequential}).
To investigate if LLMs can be used as effective sequential recommenders without training, 
\citet{hou2023large} formulated sequential recommendation as conditional ranking, employing zero-shot LLM methods to reflect user preferences from past interactions and recency. \citet{wang2023zero} developed a three-step LLM prompting to summarize user preferences, while \citet{hou2023large} and \citet{liu2023chatgpt} introduced a one-shot ICL method that utilizes the previous item interactions of the target user as a demonstration. To address position bias, \citet{hou2023large} proposed to randomize the candidate item order. In this work, we explore using training data demonstrations, not just user own history, and introduce aggregated demonstration for combining relevant users.


\section{What Makes In-Context Learning Work for Sequential Recommendation}

In this section, we conduct a preliminary empirical study to investigate the role of various aspects of demonstrations. These aspects include the wording of prompts, task consistency between demonstrations and test instances, selection of demonstrations, and number of demonstrations. 
While previous studies have explored the use of LLM as sequential recommenders in a zero-shot manner~\cite{hou2023large, wang2023zero}, this is the first study to comprehensively discuss how in-context learning can improve sequential recommendation.

\begin{table}[t]
    \caption{Dataset statistics after removing duplicate interactions and users or items with fewer than 5 interactions.}
    \vspace{-5pt}
    \setlength{\tabcolsep}{2pt}
    \resizebox{0.49\textwidth}{!}{
    \begin{tabular}{lccc}
    \toprule
    Datasets & ML-1M  & LastFM-2K & Games \\
    \midrule
    \# Users &  6,040 & 1,143 & 50,547 \\
    \# Items & 3,706 & 11,854 &  16,859\\
    \# User-item Interactions &  1,000,209 & 68,436 & 389,718\\
    Avg. interacted items per user &  165.59 & 59.92 & 7.71\\
    Avg. interacted users per item & 269.88 & 5.77 & 23.11 \\
    \bottomrule
    \end{tabular}
    }
    \label{tab:datasets}
    \vspace{-5pt}
\end{table}

\subsection{Experiment Setup}
\label{sec:analysis_expt}
We implement zero-shot, one-shot, and few-shot methods in this study, using three widely used recommendation datasets: the movie rating dataset \textit{MovieLens-1M} (ML-1M) dataset, the category of \textit{Games} from the Amazon Review dataset~\cite{mcauley2015image}, and  the music artist listening dataset \textit{LastFM-2K}~\cite{Cantador:RecSys2011}. The data statistics are summarized in Table~\ref{tab:datasets}.
Taking into account cost-effectiveness of LLMs, we select 50 data examples from each of the three datasets to carry out all experiments for analysis in Section 3.
Following the previous works~\cite{hou2023large, wang2023zero}, we use a leave-one-out strategy for evaluation, i.e., predicting the last interacted item of each user sequence and using the earlier interacted items as input.
For each user sequence, we remove the last item, keeping it aside for testing. The rest of the sequence is used for training and validation. 
To evaluate the ranking results for each user $u_i$ over a set of candidate items $c_i$, we adopt the widely used \textbf{NDCG@N} (N = 10, 20) as the evaluation metric.
For MovieLens-1M and Games, we directly use the candidate sets utilized in an earlier work~\cite{hou2023large}. For LastFM, we follow~\cite{hou2023large} and randomly select candidate items from the item universal set for each user sequence. We then insert the ground truth next item into the candidate item set.
We use ChatGPT (\texttt{GPT-3.5-Turbo}) as the default LLM due to its excellent performance and cost-effectiveness. To ensure the reliability of findings, we repeat each experiment 9 times and report the average results. Without exception, we use ML-1M as an example for discussion.

\subsection{In-Context Learning for Sequential Recommendation}

In ICL for sequential recommendation, one or a few training users are used as demonstrations that are included in the LLM prompt.  Each demonstration thus includes a training user $i$'s historical item interactions $x_i$, a set of candidates $c_i$, and ground truth next-item $y_i$. We denote the prompt capturing the demonstration user $i$ by ${\cal{T}}(x_i,c_i,y_i)$.  The following shows the concatenation of $n$ demonstrations $\cal{C}$ which is appended by the instruction prompt for the test user ${\cal{T}}(x_{test},y_{test})$ for prediction.
\begin{equation}
\mathcal{C}= \mathcal{T}\left(x_1, c_1, y_1\right)\oplus \cdots \oplus\mathcal{T}\left(x_n, c_n, y_n\right)
\end{equation}
\begin{equation}
y_{\text {test }} \sim \mathcal{P}_{LLM}\left(\cdot \mid \mathcal{C} \oplus \mathcal{T}\left(x_{\mathrm{test}}, c_{\mathrm{test}, \cdot}\right)\right)
\end{equation}

\subsection{Wording of Instructions}

\begin{figure}[t]
    \centering
    \includegraphics[width=0.46\textwidth]{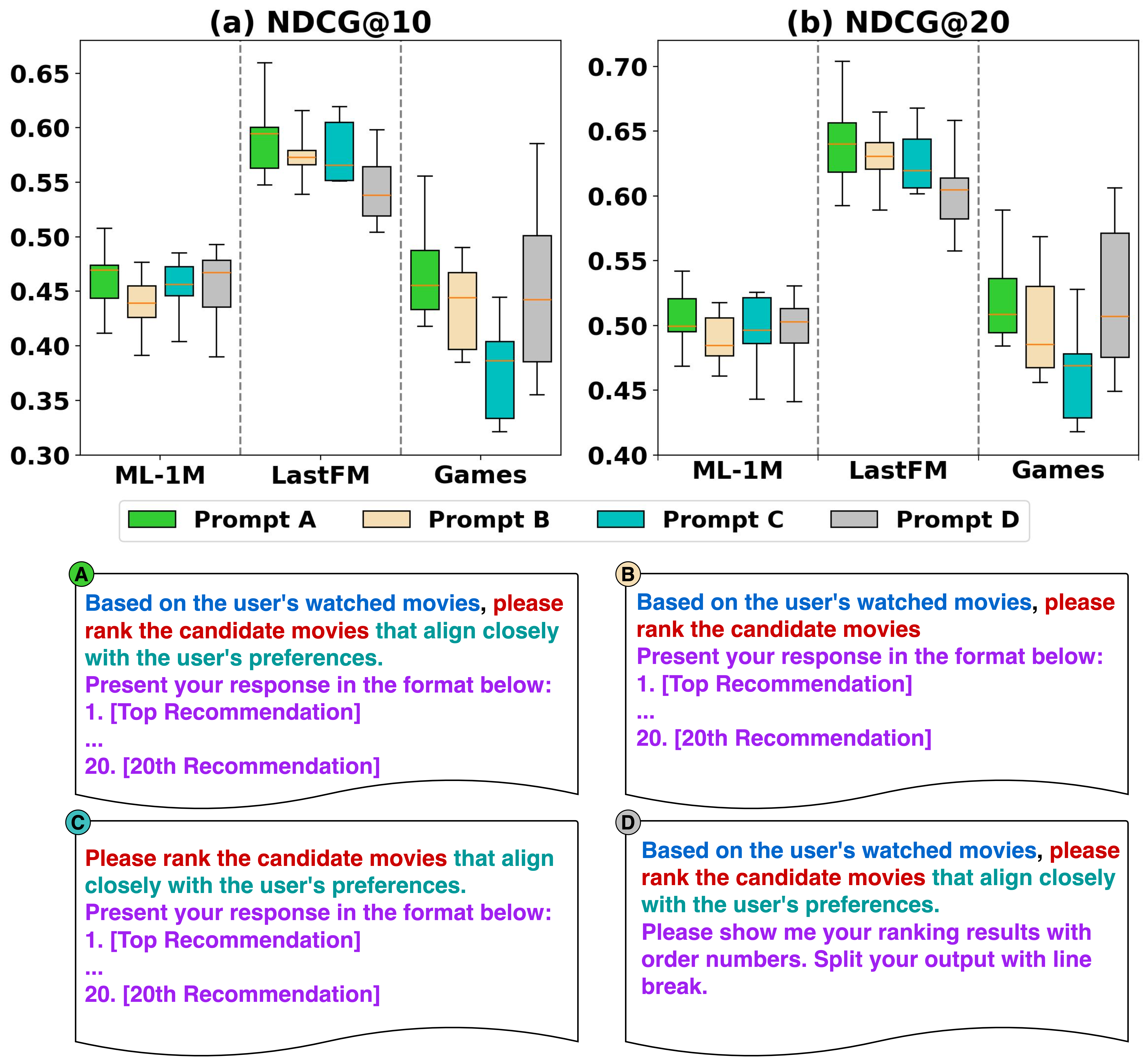}
    \caption{Instruction Format options: (A) Full, (B) w/o preference alignment, (C) w/o watched movie focus, (D) w/o rank result format}
    \label{fig:wrod_prompt}
\end{figure}

LLMs have been found to be sensitive to wording of the prompt~\cite{madaan2022text, yang2023large}. For example, prompts (or instructions) that are semantically similar may yield significantly different results~\cite{kojima2022large, zhou2022large, wang2023plan, zhang2022tempera}. 
To examine the impact of instruction wording and exclude the influence of other factors such as demonstration labels and selection, we employ LLM as a zero-shot solver for sequential recommendation.

We discuss four different options for the instruction format
to investigate the sensitivity of the LLM to the wording of the instruction.
Considering the prompts used in LLM-based zero-shot recommendation models~\cite{hou2023large, wang2023zero}, we derive instructions with four possible mention components: 
{\color{Maroon} (a) candidate item ranking}, {\color{Emerald} (b) user preference alignment}, {\color{RoyalBlue} (c) historical interacted items}, and {\color{Purple} (d) ranked result format}. 
As recommendation is formulated as a ranking task, component (b) is mandatory.  The \textit{full} instruction covers all four components. To explore better instructions, we derive other instruction options by leaving out one of the remaining components. We thus have four instruction options: (A) full instruction $\mathcal{T}^{\text{A}}$, (B) full instruction without (b) $\mathcal{T}^{\text{B}}$, (C) full instruction without (c) $\mathcal{T}^{\text{C}}$, and (D) full instruction with (d) replaced by textual result table description $\mathcal{T}^{\text{D}}$ as shown in Figure~\ref{fig:wrod_prompt}.

As shown in Figure~\ref{fig:wrod_prompt}, we observe that ChatGPT's performance degrades when the instruction does not make reference to interacted items or user preferences across three datasets. This suggests that explicit inclusion of watched movies or user preferences can improve its ability to leverage the user's historical items effectively. 
While Instruction (A) shows similar average performance as Instruction (D) on ML-1M and LastFM, the former enjoys a smaller variance and outperforms the latter on LastFm. This suggests that LLM prefers explicit output formats over textual description of output format.

\begin{tcolorbox}
    \textbf{\textit{Finding} 1.} For sequential recommendation, ChatGPT prefers explicit mentions of instructions and explicit mentions of interacted items, user preference alignment and ranked result format. 
\end{tcolorbox}

\subsection{Task Consistency}

\begin{figure}[t]
    \centering
    \includegraphics[width=0.46\textwidth]{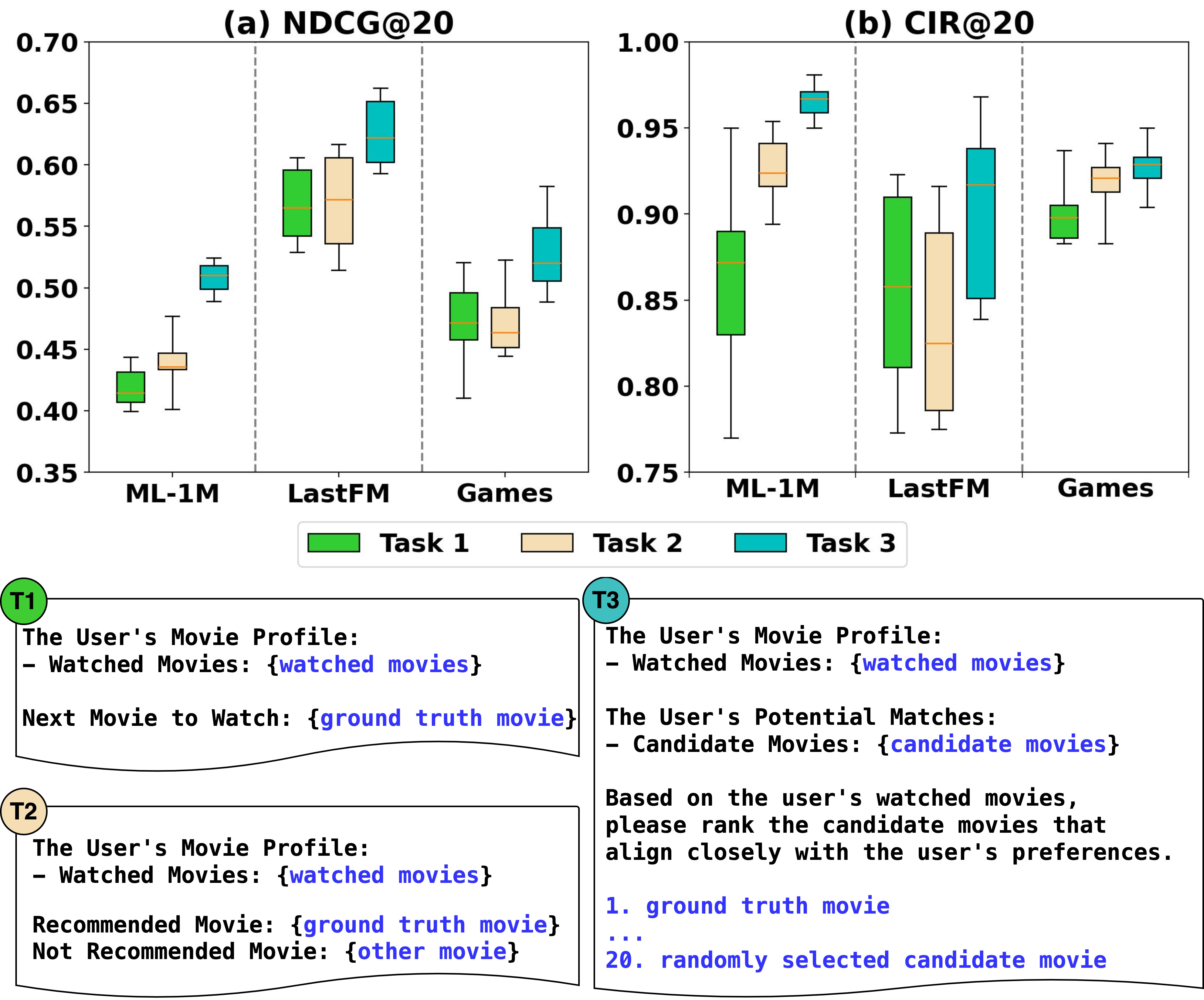}
    \caption{Impact of task consistency between demonstrations and test instances. CIR: Candidate Inclusion Ratio of Demonstration Templates: (T1) Next-Item option; (T2) Contrasting Item Pair option; (T3) Ranked Items option.}
    \label{fig:task-consistency}
\end{figure}

LLMs are capable of learning new tasks at test time by understanding the relationship between the input of a demonstration and its corresponding output label~\cite{yoo2022ground, wei2023larger}.  In sequential recommendation, LLM is required to rank the ground truth target item at the top followed by other candidate items.  However, in a demonstration example from the training set, we observe only one labeled next item but not the ranking of other candidate items.
Hence, when constructing demonstrations for in-context learning, we have to answer the important questions: How to prepare the input-label correspondence for a demonstration to be consistent with the sequential recommendation task?
To eliminate other factors that may influence the results, such as the number of demonstrations and instructions, we employ instruction (A) as it has proven to be the most effective and robust across three datasets in our previous experiments. We randomly select only one demonstration example for all experiments in this study.

In traditional sequential recommendation, next-item prediction~\cite{song2021next, petrov2022effective}, positive and negative item comparison~\cite{rendle2012bpr, kang2018self, xie2020contrastive}, and reranking~\cite{xu2023multi} are commonly utilized objectives to train models.
Hence, we develop three different prediction tasks for demonstrations for in-context learning. These tasks include: (T1) predicting the next item, 
(T2) contrasting item pairs,
and (T3) ranking candidate items. 
The prompts corresponding to these prediction tasks are shown in Figure~\ref{fig:task-consistency}.  
T1 uses the ground truth next-item directly in the demonstration. T2 uses the ground truth next item and another randomly selected item as the positive and negative items respectively. T3 ranks the ground truth next item at the first position and randomly shuffles the remaining candidate items to fill the other positions.
Among the task prediction task options, T3 is the only one that aligns closely with the instruction for the test user, i.e., (A).

Figure~\ref{fig:task-consistency} shows the results of these three tasks across three datasets. T3 consistently outperforms T1 and T2 on all three datasets, suggesting that task consistency between demonstration and test user  benefits in-context learning for sequential recommendation. 
Additionally, 
As the recommended items may not be found among the provided candidates, we also report \textit{candidate inclusion ratio} (CIR) which measures the proportion of the candidate items that appear in the ranked item results. As shown in Figure~\ref{fig:task-consistency}, we observe that the CIR generally correlates with the NDCG results. The inconsistent demonstration task options (e.g., T1 and T2 coupled with test instruction option (A)) are more likely to cause the LLM to generate non-candidate items in the results. This helps to understand why T3 achieves the best performance.
\begin{tcolorbox}
    \textbf{\textit{Finding} 2.} Maintaining task consistency between demonstrations and test users is beneficial for in-context learning in sequential recommendation.
\end{tcolorbox}

\subsection{Selection of Demonstrations}

\begin{figure}[t]
    \centering
    \includegraphics[width=0.48\textwidth]{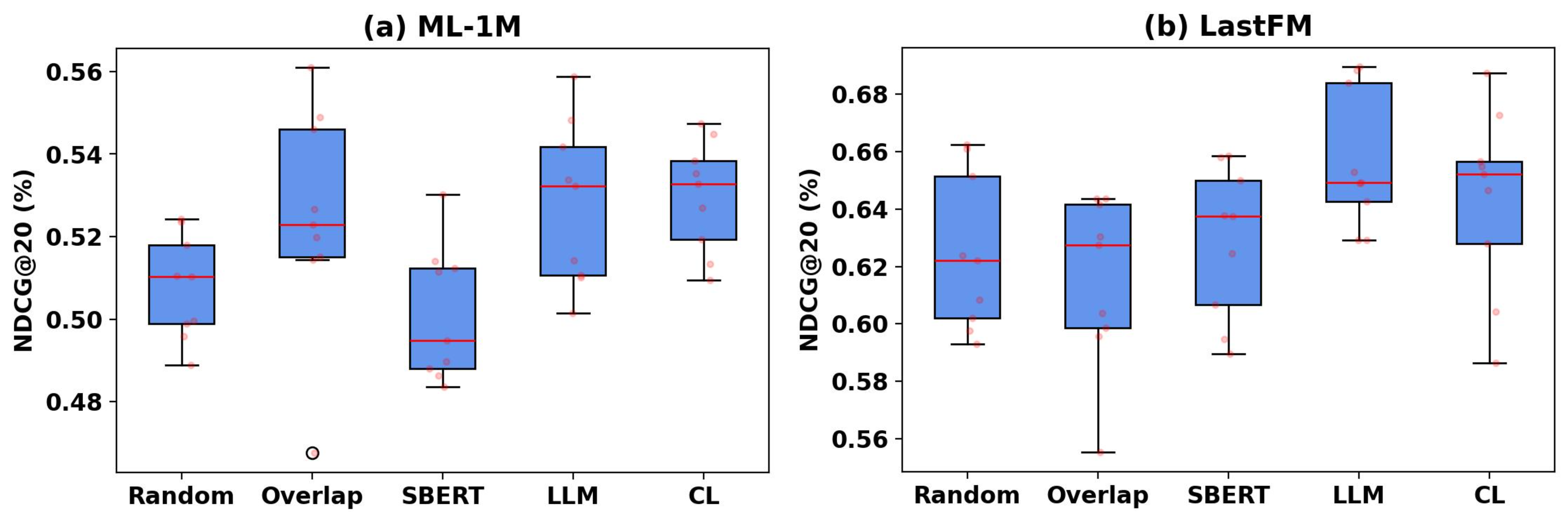}
    \caption{Demonstration selection: (1) random selection; (2) overlapping interacted items; (3) cosine similarity between the SBERT embeddings of interacted item sequences; (4) cosine similarity between the LLM (OpenAI embeddings) of interacted item sequences; (5) cosine similarity using CL embeddings of interacted item sequences.}
    \label{fig:selection-demonstration}
\end{figure}

It has been observed that the performance of in-context learning greatly depends on selecting suitable demonstrations~\cite{liu2021makes}. Utilizing examples that are semantically similar to the test sample can provide more informative and task-relevant knowledge to LLMs. Following ~\citet{liu2021makes}, there are several follow-up works~\cite{rubin2021learning, shi2022xricl, zhang2023makes, li2023unified} to develop methods for selecting better demonstrations.
In this work, we evaluate five different demonstration selection methods to determine their impact to in-context learning for sequential recommendation. These methods include: (1) random selection; (2) overlapping historical items of demonstration user and test user; (3) text similarity scores using Sentence-BERT embedding~\cite{reimers2019sentence} (SBERT); (4) text similarity scores using LLM OpenAI embedding\footnote{\texttt{text-embedding-ada-002} (https://platform.openai.com/docs/models/moderation)} (LLM); and (5) trained retriever using contrastive learning~\cite{xie2020contrastive, li2023unified} (CL). 
In Option (5), positive examples are obtained by data augmentation applied to the anchor user sequence, while negative examples are randomly selected user item-interaction sequences.

Figure~\ref{fig:selection-demonstration} compares the five selection methods on ML-1M and LastFM as they are used in one-shot sequential recommendation. 
The results show that selection methods (4) and (5) generally outperform the rest. As method (4) appears to be more robust than (5) and it does not require additional training, 
we thus use that as the default retriever model in the subsequent experiments. 

\begin{tcolorbox}
    \textbf{\textit{Finding} 3.} Retrieval-based methods are better than random selection, and stronger LLMs can serve as stronger retrievers without any training.
\end{tcolorbox}

\subsection{Number of Demonstrations}
\label{sec:analysis_demonstrations}

\begin{figure}[t]
    \centering
    \includegraphics[width=0.48\textwidth]{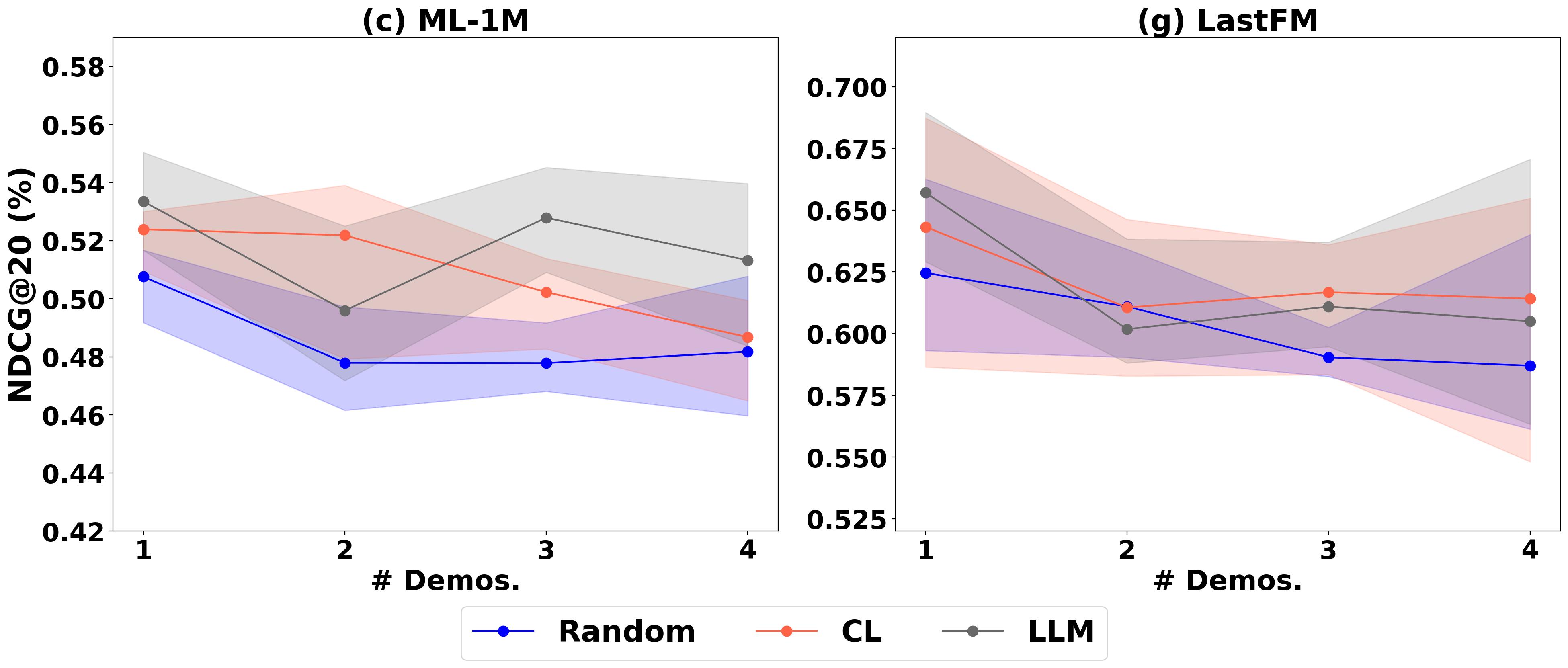}
    \caption{Varying number of demonstrations.}
    \label{fig:number-demo}
\end{figure}

When training a model, having more training data examples usually leads to better model performance. However, it is interesting to note that ~\citet{zhao2023dynamic} discover that increasing the number of demonstrations for in-context learning does not necessarily result in improved performance. Similarly, ~\citet{chen2023many} finds that using only one demonstration may not perform worse than using more demonstrations. In our case, we evaluate the impact of the number of demonstrations on ML-1M and LastFM using random selection, LLM, and CL demonstration selection methods. We conduct experiments with the number of demonstrations ranging from 1 to 4, as exceeding 4 demonstrations would exceed the input limit of ChatGPT (\texttt{GPT-3.5-Turbo}). Figure~\ref{fig:number-demo} demonstrates a clear trend of performance decreasing with number of demonstrations.
\begin{tcolorbox}
    \textbf{\textit{Finding} 4.} Increasing demonstrations for in-context learning for sequential recommendation would result in performance degradation and exceed the input limit of LLMs.
\end{tcolorbox}

\section{In-Context Learning with Aggregated Demonstrations}
\label{sec:iCL_syn_user}

Finding 4 suggests LLMs have difficulties coping with multiple demonstrations in sequential recommendation. 
A similar finding by \citet{liu2023lost} also suggests that the current language models often struggle to utilize information in long input contexts. In particular, their performance tends to significantly degrade when the relevant information is located in the middle of long contexts, also known as the ``lost in the middle'' phenomenon. 
The in-context learning prompts for sequential recommendation can easily exceed the prompt length limit of LLM when more than 4 demonstrations are to be accommodated.  Such prompts not only suffer from ``lost in the Middle'', but also incur additional costs of calling LLM APIs.

To address the above challenge, we propose {\it aggregated demonstration} which combines $K$ ($K>1$) demonstration users into one for in-context learning.
This simple yet effective in-context learning method for sequential recommendation is called \textbf{LLMSRec-Syn}. As the prompt length of aggregated demonstration only increases marginally when we increase $K$, LLMSRec-Syn can accommodate more member demonstration users.

Based on Finding 3, LLMSRec-Syn begins with selecting $K$ demonstration users that are similar to the test user.  We use similarity between the LLM embeddings of demonstration and test users. 
We also follow Finding 1 and adopt instruction template (A) for the test user. Based on Finding 2, we also adopt demonstrate template (T3) for the aggregated demonstration to maintain consistency with the task for test user.
Next, we construct the aggregated demonstration's historical item-interactions, candidate items, and the desired ranking of the candidate items from its member demonstrations, as shown in Figure~\ref{ch5_fig:aggregated_prompt} in the Appendix.

\noindent\textbf{Historical item-interactions}. Let $H$ denote the historical item-interactions and $H$ is empty initially. We first rank the $K$ selected demonstration users by similarity score.  We then add the most recent interacted item from the most similar demonstration to $H$.
We repeat the same step for the remaining demonstrations in their similarity order. 
When we run out of most recent interacted items from $K$ selected demonstrations, we continue to add the next recent interacted items of these demonstrations to $H$ until the number of historical items reaches $MAX_H$. 

\noindent\textbf{Candidate Items}. Let $C$ denote the candidate items of the aggregated demonstration and $C$ is empty initially.  We first gather all the ground truth next items from the $K$ selected demonstrations and add them to $C$. Next, we randomly add other items from the item pool to $C$ so as to meet the required number of candidate items. 

\noindent\textbf{Ranking of Candidate Items}. To rank the candidate items in $C$, we place the ground truth next item of the most similar demonstration at rank 1, followed by that of next similar demonstration until we run out of the ground truth next items of all $K$ selected demonstrations.  Next, we assign random ranks to the remaining items in $C$. 

Once the aggregated demonstration is constructed, it is added to the prompt the same way a training user is added as a demonstration.   we add it to the corresponding test user and use them as input for the LLM. 
\begin{equation}
    \begin{split}
        \mathcal{C}^{\mathbf{a g g}}=\mathcal{T}^{\mathrm{A}}(\operatorname{Agg}^{\mathrm{T} 3}(x_{\sigma_1},   &c_{\sigma_1}, y_{\sigma_1}, \cdots, x_{\sigma_n}, \\ c_{\sigma_n}, y_{\sigma_n})),
    \end{split}
\end{equation}
\begin{equation}
y_{\text {test }} \sim \mathcal{P}_{L L M}\left(\cdot \mid \mathcal{C}^{\mathbf{agg}} \oplus \mathcal{T}^{\text{A}}\left(x_{\text {test }}, c_{\text {test }}\right)\right),
\end{equation}
where $\sigma_i$ represents the $i^{\rm th}$ ranked selected users returned by the retrieval model.
Finally, the LLM generates a ranked list of candidate items as the recommendation result.

There are several advantages of the proposed LLMSRec-Syn: 1) Standard demonstration only has one ground truth next item in the ranking list. In contrast, the aggregated demonstration includes more next items at high positions in the ranking list. This approach can avoid sparse signals and provide more guidance to LLMs for recommending to the test user;
2) LLMSRec-Syn is less sensitive to the number of demonstrations;
3) Cost of LLMSRec-Syn does not increase much with the number of demonstrations; and
4) LLMSRec-Syn keeps to the prompt length limit of LLMs.

\section{Experiments and Results}

\subsection{Methods for Comparison}
To evaluate the performance of LLMSRec-Syn, we conduct an extensive set of experiments on ML-1M, Games, and LastFM-2K datasets.  
Following~\citet{hou2023large}, we select 200 data examples from each of the three datasets to carry out all experiments.
We use an experiment setup similar to that mentioned in Section~\ref{sec:analysis_expt} except that we now uses more LLMs and reports the NDCG@N results where N=5,10, and 20.
We compare LLMSRec-Syn with 10 methods categorized into 3 types: 

\noindent\textbf{Supervised methods}: Most Popular (Recommending items based on their overall popularity among all users in the training data), GRU4Rec~\cite{hidasi2015session} (using GRUs to model user's item sequences), and SASRec~\cite{kang2018self} (employing a self-attention mechanism to learn user preferences from their item sequences). 

\noindent\textbf{Zero-shot methods}:
BM25~\cite{robertson2009probabilistic} (ranking candidate items based on their textual similarity with the test user's interacted items), LLMSeqSim~\cite{harte2023leveraging} (ranking candidate items by semantically similarity using OpenAI embeddings (\texttt{text-embedding-ada-002})), LLMRank-Seq~\cite{hou2023large} (using ChatGPT to rank candidate items with crafted prompts),
and LLMSRec (a zero-shot version of the proposed LLMSRec-Syn using the instruction prompt $\mathcal{T}^{\text{A}}$).

\noindent\textbf{One-shot methods}: LLMRank-His~\cite{hou2023large} (using historical items of the test user to form a demonstration), LLMSRec-Fixed (using a randomly selected demonstration for all test users), and LLMSRec-Nearest (finding the most similar training user as the demonstration).

As Section~\ref{sec:analysis_demonstrations} shows that more than one demonstration in in-context learning for sequential recommendation does not yield better performance, we do not include few-shot methods in this set of experiments. We however will study how many member demonstrations $K$ is ideal for aggregated demonstration (see Section~\ref{sec:main_exp_result}). 

We implement LLMSRec-Syn using three different LLMs, LLaMa2~\cite{touvron2023llama}, ChatGPT~\cite{openai-chatgpt-2022} (LLMSRec-Syn), and GPT-4~\cite{openai-gpt4-2023} (LLMSRec-Syn-4).
For the LLMSRec-Syn-4 experiment, which is shown in the last row of Table~\ref{tab:main_results}, we used GPT-4 as the base LLM. For all other experiments, including preliminary studies, in-depth analysis, and method comparisons presented in Table~\ref{tab:main_results}, we used the same ChatGPT (\texttt{GPT-3.5-Turbo}). To ensure the reliability of our findings, each experiment is conducted 9 times, and the average results are reported.
However, we found LLaMa2 unable to follow recommendation instructions and is prone to generating historical interacted items or in-context examples. As a result, we exclude the LLaMa2 results
In LLMSRec-Syn, we set the number of member users in the aggregated demonstration as \{1,2,3,4,5,6,7\} and conduct a brute force search to determine the optimal number for each dataset. 
We set the number of historical items $MAX_H=50$ and number of candidate items to 20. We analyse some specific test cases of LLMSRec-Syn-4 in the Appendix~\ref{appendix:case}.

\subsection{Main Results}
\label{sec:main_exp_result}

\begin{table*}[!htp]\centering


\caption{Main results. We report 
NDCG@5, NDCG@10 and NDCG@20 on ML-1M, LastFM-2K and Games. (Best results in each group of methods are \textbf{boldfaced} and overall best results are \underline{underlined}).
}
 \vspace{-5pt}
\resizebox{1.0\textwidth}{!}{ 
\setlength{\tabcolsep}{2.2pt}
\begin{tabular}{ll|ccc|ccc|ccc}\toprule
\multirow{ 2}{*}{Setting}& \multirow{ 2}{*}{Method} & \multicolumn{3}{c|}{ML-1M} & \multicolumn{3}{c|}{LastFM-2K} &  \multicolumn{3}{c}{Games} \\
 & &NDCG@5 &NDCG@10 &NDCG@20 &NDCG@5 &NDCG@10 &NDCG@20 &NDCG@5 &NDCG@10 &NDCG@20\\
 \midrule

\multirow{ 3}{*}{Supervised}
& Most Popular  & {0.3673}  &  {0.4623} & {0.4748} & {0.4055} & {0.4205} & {0.4803} & {0.2746} & {0.3905} & {0.4496}\\
& GRU4Rec  & {0.7205}  &  {0.7494} & {0.7610} & {0.3382} & {0.3971} & {0.4784} & {0.6747} & {0.7002} & {0.7278}\\
& SASRec  & \underline{{\bf 0.7322}} & {\underline{\bf 0.7595}} &  {\underline{\bf 0.7702}} &  {\bf 0.4081}  & {\bf 0.4680} & {\bf 0.5303} & \underline{{\bf 0.6828}} & \underline{{\bf 0.7189}}  & \underline{{\bf 0.7311}}\\


\midrule

\multirow{ 4}{*}{Zero-shot}
& BM25  &0.1314 & 0.2053 &0.3370  &0.1215 & 0.1393 & 0.3354 & 0.2285 & 0.3108  & 0.4055\\



& LLMSeqSim   & {0.3250}  &  {0.4037} & \bf {0.4723} & {0.4090} & {0.4662} & {0.5293} & \bf{0.4269} & \bf{0.4830} & \bf{0.5360}\\

 & LLMRank-Seq 
& \bf 0.3344 & 0.3882 & 0.4612 
& 0.5084 &  0.5545  &\bf 0.6070 & 0.3063 & 0.3607 & 0.4074 \\
& LLMSRec   & {0.3339}  & \bf {0.4087} & \bf{0.4723} & \bf {0.5126} & \bf {0.5602} & {0.6057} & {0.4070} & {0.4555} & {0.5103}\\

\midrule


\multirow{ 5}{*}{ One-shot}
& LLMRank-His   & {0.3919}  &  0.4444 & 0.5074 & {0.5318} & {0.5725} & {0.6212} & {0.4191} & {0.4667} & {0.5206}\\

& LLMSRec-Fixed 
& 0.3590 & 0.4193 & 0.4793 
& 0.4961 & 0.5425 & 0.5984 & 0.3744 & 0.4400 & 0.4899 \\


& LLMSRec-Nearest
&0.3842 & 0.4382 & 0.5017
& 0.5249 & 0.5697 & 0.6197 & 0.3975 & 0.4388 & 0.4994 \\

& LLMSRec-Syn
&  0.4267 & 0.4813 & 0.5334 
& 0.5554 & 0.5918 & 0.6371 & 0.4989 & 0.5334 & 0.5869 \\
& LLMSRec-Syn-4
&  {\bf 0.5112} & {\bf 0.5685} & {\bf 0.5936} 
& {\underline{\bf 0.6544}} & {\underline{\bf 0.6799}} & {\underline{\bf 0.7017}} & {\bf 0.5647} & {\bf 0.6019} & {\bf 0.6277} \\


\bottomrule
\end{tabular}
}
\label{tab:main_results}
 \vspace{-5pt}
\end{table*}

The main experiment results are shown in Table~\ref{tab:main_results}, from which we obtain the following findings:

\noindent\textbf{ICL one-shot methods with appropriate demonstrations out-perform zero-shot methods}. As shown in Table~\ref{tab:main_results}, LLMRank-His, LLMSRec-Fixed, and LLMSRec-Nearest using one training user as demonstration outperform LLMRank-Seq on three datasets, except for LLMSRec-Fixed which performs slightly worse than LLMRank-Seq on LastFM-2K.  This result suggests that ICL can enhance the LLM's ability to perform a complex task such as sequential recommendation.

\noindent\textbf{Aggregated demonstration, combining multiple member users, allows LLM to effectively gather useful task specific information about the test user within a concise context}. Compared to other ICL baselines (i.e., LLMRank-His, LLMSRec-Fixed, and LLMSRec-Nearest), LLMSRec-Syn achieves the superior one-shot performance across all datasets as shown in Table~\ref{tab:main_results}.  While Figure~\ref{fig:number-demo} shows that having more demonstrations may hurt ICL for sequential recommendation, the idea of incorporating multiple demonstration users into an aggregated demonstration enhances the performance of LLMSRec-Syn. These results illustrate the advantage of aggregated demonstration in accommodating multiple training users within a limited prompt length.

\noindent\textbf{LLMSRec-Syn is competitive against supervised methods when the amount of training data is limited}. LLMSRec-Syn easily outperforms the simple supervised baseline, Most Popular.  While it does not outperform GRU4Rec and SASRec on ML-1M and Games, LLMSRec-Syn surprisingly outperforms all supervised baselines on LastFM-2K.  One possible reason is that LastFM-2K has sparse information about items after removing duplicate user-item interactions and users/items with less than 5 interactions, making it challenging to train a good supervised model. 

\noindent\textbf{LLMSRec-Syn using more powerful LLMs may outperform supervised methods in the future}. With rapid advancement of LLM research, LLMSRec-Syn can be further enhanced when more powerful LLM is used. Our results in Table~\ref{tab:main_results} shows that LLMSRec-Syn-4 significantly outperforms LLMSRec-Syn on all the 3 datasets.

\subsection{Analysis of Aggregated Demonstrations}

In this section, we study the recommendation performance when varying the settings of aggregated demonstrations. Analysis of ordering of users and label in the aggregated demonstration can be found in the Appendix~\ref{appendix:analysis}.

\begin{figure}[t]
    \centering
    \includegraphics[width=0.48\textwidth]{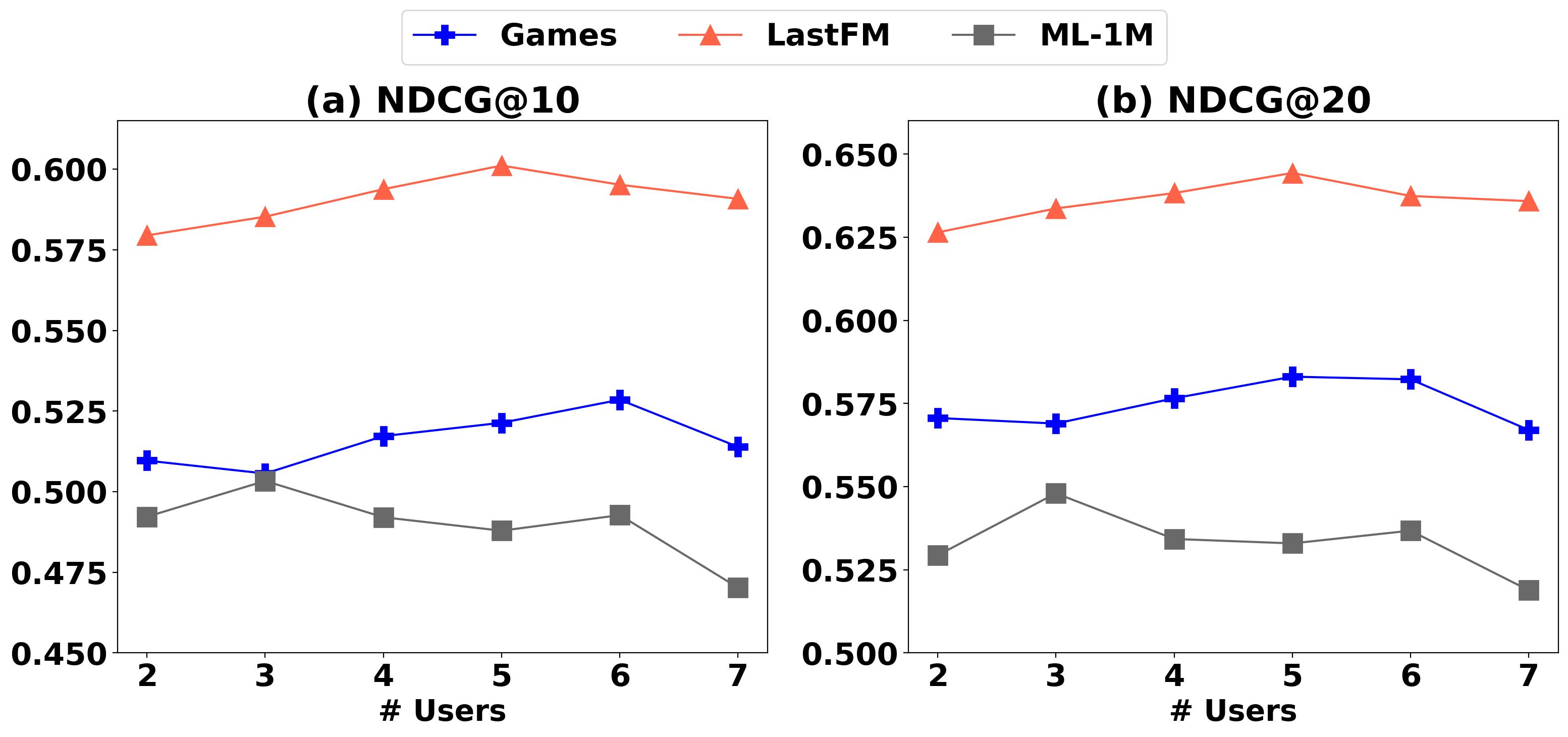}
    \vspace{-5pt}
    \caption{Varying number of users ($K$) in aggregated demonstration.}
    \label{fig:number-user-agg}
    \vspace{-5pt}
\end{figure}

\noindent\textbf{Impact of number of users in the aggregated demonstration}. We evaluate the impact of $K$ (the number of member users) in the aggregated demonstration on LLMSRec-Syn's performance.  We empirically vary $K$ from 2 to 7. As shown in Figure~\ref{fig:number-user-agg}, an approximate inverted U-shaped relationship exists between $K$ and NDCG@10/20 performance. Initially, as $K$ increases, there is a noticeable performance increase, suggesting that LLMSRec-Syn benefits from aggregated demonstration. However, beyond some $K$ value, more member users in aggregated demonstration leads to lower performance. This can be explained by more irrelevant training users being incorporated into the aggregated demonstration.

\begin{figure}[t]
    \centering
    \includegraphics[width=0.48\textwidth]{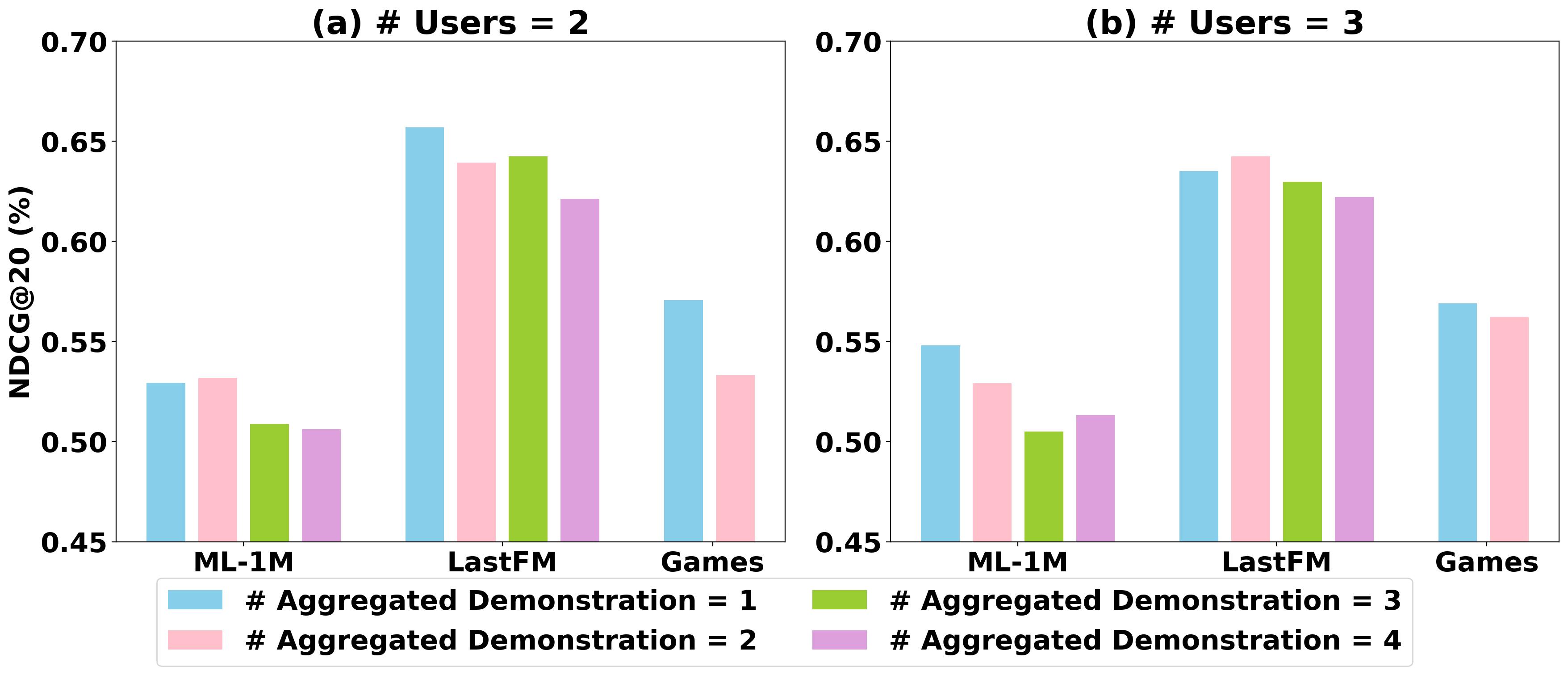}
    \vspace{-5pt}
    \caption{Varying number of aggregated demonstrations each with: (a) 2 member users, and (b) 3 member users.}
    \label{fig:vary-number-agg}
    \vspace{-5pt}
\end{figure}

\noindent\textbf{Impact of number of aggregated demonstrations.} We evaluate the impact of the number of aggregated demonstrations to LLMSRec-Syn by varying the number of aggregated demonstrations from 1 to 4 such that each demonstration involves 2 users (see Figure~\ref{fig:vary-number-agg}(a)) and 3 users (see Figure~\ref{fig:vary-number-agg}(b)). 
For the Games dataset, experimentation with 3 aggregated demonstrations was not possible due to GPT-3.5-Turbo's input limit. The results show that a single aggregated demonstration outperforms multiple ones, except in the LastFM-2K dataset, where two demonstrations slightly excel.

\section{Conclusion}
This paper investigates in-context learning (ICL) for LLM-based sequential recommendation. Our study identifies key factors such as instruction format and demonstration selection that influence ICL's effectiveness. We further introduce the LLMSRec-Syn method which utilizes our proposed aggregated demonstration to efficiently incorporate relevant information from multiple training users. Tested on three datasets, LLMSRec-Syn consistently outperforms existing LLM-based sequential recommendation methods. Future work includes a detailed analysis of LLMSRec-Syn's unexpected success compared to some supervised methods and the optimization of aggregated demonstration strategies.

\section{Limitations}
While this paper considers several factors in applying LLMs to sequential recommendation and proposes a new demonstration concept known as aggregated demonstration, there are still some limitations yet to be addressed.
Firstly, the wording of LLMSRec-Syn prompt is manually handcrafted and may not be optimal.
This concern is also mentioned in works on prompt optimization~\cite{yang2023large, deng2022rlprompt, pryzant2023automatic}. However, determining the optimal prompt wording typically requires feedback (such as validation set results~\cite{yang2023large}), carefully designed reward function~\cite{deng2022rlprompt}, or textual feedback from large language models to iteratively update the initial prompt~\cite{pryzant2023automatic}.
Moreover, when a user's historical items are too many, LLMSRec-Syn may still suffer from the issue of long text.
Furthermore, the aggregated demonstration method, while mitigating input length constraints, might oversimplify the user preferences, potentially resulting in less personalized recommendations.
Moreover, the non-utilization of existing user datasets for pretraining or fine-tuning, due to LLM size constraints, limits the adaptability and fine-tuning of the model to specific recommendation contexts.
These limitations highlight the need for further research in optimizing LLMs for complex, dynamic tasks such as sequential recommendation, where user context and historical data play crucial roles.

\bibliography{custom}

\clearpage
\section{Appendix}

\subsection{More In-Depth Analysis}
\label{appendix:analysis}
\begin{figure}[t]
    \centering
    \includegraphics[width=0.48\textwidth]{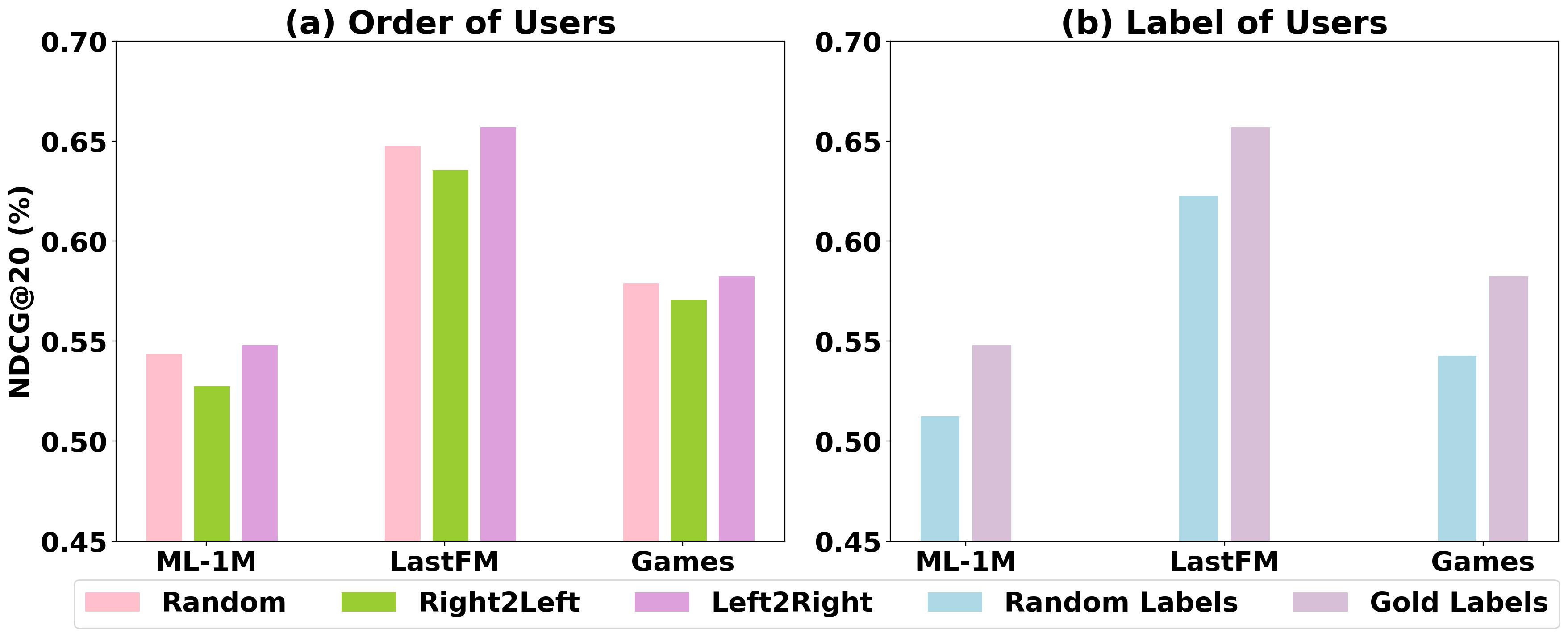}
    \caption{(a) Ordering of member users in the aggregated demonstration. (b) Ground truth vs random next-items in aggregated demonstrations.}
    \label{fig:sec53}
\end{figure}




\begin{figure*}[t]
    \centering
    \includegraphics[width=0.9\textwidth]{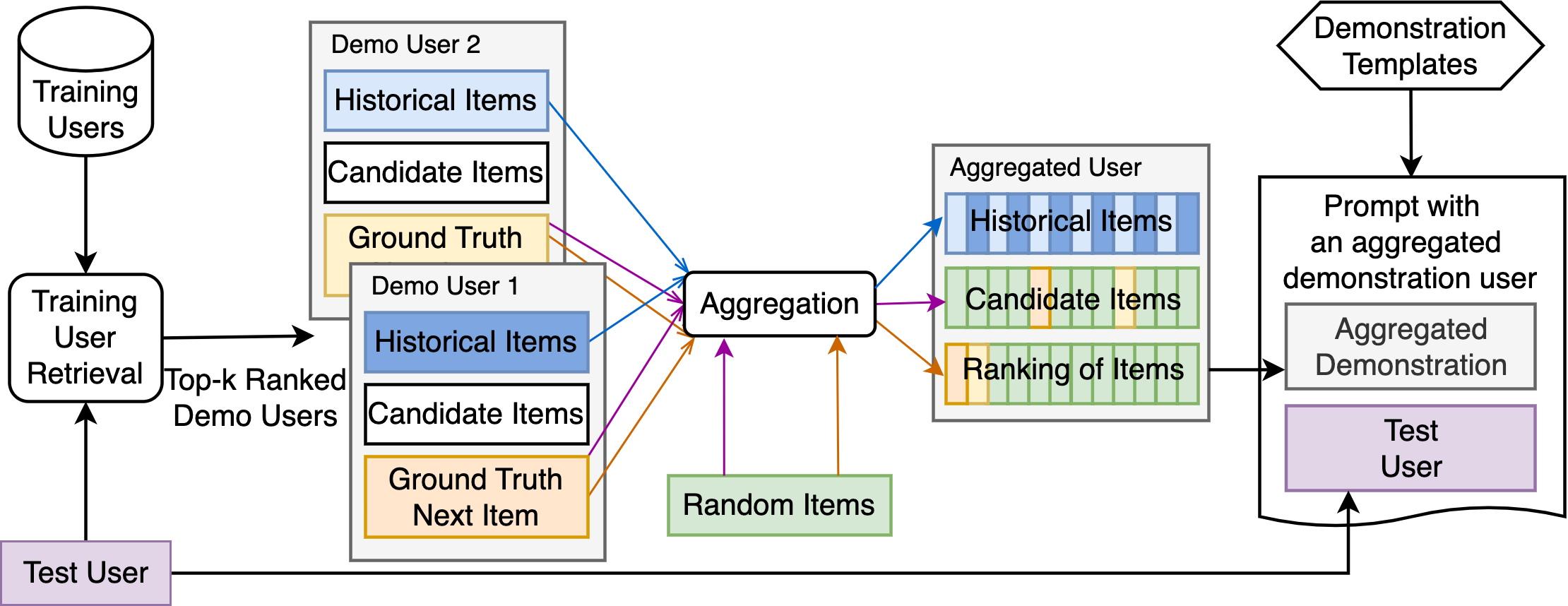}
    \caption{Construction of in-context learning prompt with aggregated demonstration for sequential recommendation.}
    \label{ch5_fig:aggregated_prompt}
    \vspace{-5pt}
\end{figure*}

\noindent\textbf{Impact of user order in aggregated demonstrations}. We experiment with 3 possible orders of member users: (i) Random (randomly selects historical items and next-items from the selected users to construct the aggregated demonstration), (ii) Right2Left (the reverse order of demonstration users in constructing an aggregated demonstration in LLMSRec-Syn in Section~\ref{sec:iCL_syn_user}), and (iii) Left2Right (the user order used in the LLMSRec-Syn). Figure~\ref{fig:sec53}(a) illustrates that Left2Right and Right2Left are the most and least ideal orders respectively. The performance of Random is naturally sandwiched in between.

\noindent\textbf{Impact of labeled next-items in the aggregated demonstration}.
According to~\citet{min2022rethinking}, ground truth labels are not important for in-context learning. To investigate this claim for ICL-based sequential recommendation, we compare LLMSRec-Syn using ground truth next-items in the aggregated demonstration (referred to as ``Gold Labels'') with that using random non-ground truth next-items (referred to as ``Random Labels''). Our results in Figure~\ref{fig:sec53}(b) clearly indicate that ground truth next-items are required to yield better performance contradicting the claim by ~\citet{min2022rethinking}.  This could possibly be explained by the complexity of sequential recommendation task.

\begin{table}[t]
\centering
    \caption{Further comparison of different tasks with candidates. We have developed two methods with the same information as T3: (1) T1 w/ Candidate (adding candidate items in the T1 prompt) and (2) T2 w/ Candidate (adding candidate items in the T2 prompt).}
    \setlength{\tabcolsep}{1.8pt}
    \resizebox{0.49\textwidth}{!}{
    \begin{tabular}{l|ccccc}
    \toprule
    ML-1M & T1  & T1 w/ Cand. & T2 & T2 w/ Cand. & T3 \\
    \midrule
    NDCG@10 &  0.3640 & 0.3766 & 0.3776 & 0.3972 &0.4584 \\
    NDCG@20 & 0.4193 & 0.4420 & 0.4384 & 0.4510 &  0.5077 \\
    \bottomrule
    \end{tabular}
    }
    \label{ch5_tab:temp_cand}
\end{table}

\noindent\textbf{Further comparison of different tasks with candidates.} 
As shown in Table~\ref{ch5_tab:temp_cand}, we observed that T1(T2) with candidate items in the prompt performs better than T1(T2). These results support the reviewer's comment that including more information in the prompt will enhance the performance. However, in this more fair comparison, T3 still outperforms T1 with candidate items in the prompt and T2 with candidate items in the prompt.

\begin{table}[t]
\centering
    \caption{Results of fine-tuned LLaMa2 with LoRA for in-context sequential recommendation. Regular means LLaMa2-LoRA-Regular. Aggregated means LLaMa2-LoRA-Aggregated. }
    \resizebox{0.45\textwidth}{!}{
    \begin{tabular}{l|ccc}
    \toprule
    ML-1M & Regular &	Aggregated	& LLMSRec-Syn \\
    \midrule
    NDCG@10 &  0.3640 & 0.3766 & 0.3776  \\
    NDCG@20 & 0.4193 & 0.4420 & 0.4384   \\
    \bottomrule
    \end{tabular}
    }
    \label{ch5_tab:llama2}
\end{table}
\noindent\textbf{Could fine-tuned LLaMa2 improve the performance of in-context sequential recommendation?}
We initially used a training dataset of 150 data examples to train LLaMa2 with LoRA, which we referred to as LLaMa2-LoRA-Regular. For each training data example in this training dataset, the target output is the ranking of the candidate items for a training user. The input consists of a regular demonstration example, as well as historical items and candidate items from the training user. After training, we evaluated the performance of LLaMa2-LoRA-Regular using the same 50 test users as ChatGPT-based LLMSRec-Syn (0.5283 NDCG@10). 

As shown in Table~\ref{ch5_tab:llama2}, the results showed that LLaMa2-LoRA-Regular achieved a NDCG@10 score of 0.2344. To investigate whether aggregated demonstration helps to train a better model compared to regular demonstrations, we prepared a training dataset using aggregated demonstrations instead of regular demonstrations. We trained LLaMa2 with LoRA using this dataset, which we call LLaMa2-LoRA-Aggregated. LLaMa2-LoRA-Aggregated achieved a NDCG@10 score of 0.3432 on the same test set.
Although the initial study indicates that LLaMa-LoRA performs worse than ChatGPT, the fine-tuned LLaMa2-LoRA appears to have the potential to enable in-context learning-based sequential recommendation and aggregated demonstration can help to train a better model.

\subsection{Case Study Examples}
\label{appendix:case}
In this section, we provide comparative examples of one-shot LLMSRec-Syn (Table~\ref{chap_5:case1}), one-shot LLMSRec-Nearest (Table~\ref{chap_5:case2}), one-shot LLMSRec-Fixed (Table~\ref{chap_5:case3}), and zero-shot LLMSRec (Table~\ref{chap_5:case4}). Observations show that LLMSRec-Syn ranks the ground truth movie higher than the other methods. Compared to Nearest and Fixed demonstrations, the aggregated demonstration allows the LLM to better identify a user's interests and align the ranking with those interests. Without demonstration, zero-shot LLMSRec relies solely on the LLM's knowledge and performs poorly. This suggests that LLMs can learn from demonstrations to improve in areas where they might not originally be good at.

\newpage
\onecolumn
\begin{spacing}{1.0}
\begin{center}
\ttfamily
\footnotesize
\begin{longtable}{p{0.96\linewidth}}
    \caption[Example of LLMSRec-Syn on the ML-1M dataset]{Example of the one-shot LLMSRec-Syn on the ML-1M dataset. Ground truth recommendation is highlighted in {\color{Maroon}Maroon}.} \label{chap_5:case1} \\ \toprule
    {\color{blue}\textbf{Aggregated Demonstration Example:}} \\\\
\textbf{The User's Movie Profile}:\\
- Watched Movies: {\color{violet}[`0. Caddyshack', `1. Glory', `2. A Bug's Life', `3. Star Trek VI: The Undiscovered Country', `4. Indiana Jones and the Last Crusade', `5. The Color of Money', `6. Raging Bull', `7. Edward Scissorhands', `8. Kramer Vs. Kramer', `9. Roger \& Me', `10. Romancing the Stone', `11. Full Metal Jacket', `12. The Shining', `13. Easy Rider', `14. Glory', `15. The Color Purple', `16. Die Hard', `17. Who Framed Roger Rabbit?', `18. Ghostbusters', `19. The Right Stuff', `20. No Way Out', `21. The Breakfast Club', `22. Dead Poets Society', `23. One True Thing', `24. Full Metal Jacket', `25. U2: Rattle and Hum', `26. Caddyshack', `27. Jaws', `28. Get Shorty', `29. A Fish Called Wanda', `30. Star Trek IV: The Voyage Home', `31. Brazil', `32. The Fugitive', `33. Willy Wonka and the Chocolate Factory', `34. Ghostbusters', `35. The Blues Brothers', `36. Stripes', `37. Stand by Me', `38. Pleasantville', `39. Bull Durham', `40. Risky Business', `41. Indiana Jones and the Last Crusade', `42. Star Wars: Episode VI - Return of the Jedi', `43. Witness', `44. Brazil', `45. Predator', `46. The Terminator', `47. Grosse Pointe Blank', `48. Contact', `49. The Princess Bride']}\\

\textbf{The User's Potential Matches}:\\
- Candidate Movies: {\color{violet}[`0. Excalibur', `1. The Fox and the Hound', `2. The Cable Guy', `3. Cat on a Hot Tin Roof', `4. Slingshot, The (Kådisbellan )', `5. Institute Benjamenta, or This Dream People Call Human Life', `6. Prefontaine', `7. Star Wars: Episode IV - A New Hope', `8. 8 Seconds', `9. Two Family House', `10. American Graffiti', `11. Seven Days in May', `12. The English Patient', `13. Jurassic Park', `14. Meet the Parents', `15. Weekend at Bernie"s', `16. Kurt \& Courtney', `17. Children of a Lesser God', `18. The Haunting', `19. Love and Basketball']}\\
Based on the user's watched movies, please rank the candidate movies that align closely with the user's preferences. \\
- You ONLY rank the given Candidate Movies.\\
- You DO NOT generate movies from Watched Movies.\\

Present your response in the format below:\\ 
1. [Top Recommendation (Candidate Movie)] \\ 
2. [2nd Recommendation (Candidate Movie)]\\ 
...\\
20. [20th Recommendation (Candidate Movie)]\\
\textbf{Answer}:\\
1. Children of a Lesser God\\
2. Excalibur\\
3. Jurassic Park\\
4. American Graffiti\\
5. Meet the Parents\\
6. The Haunting\\
7. Institute Benjamenta, or This Dream People Call Human Life\\
8. Cat on a Hot Tin Roof\\
9. Slingshot, The (Kådisbellan )\\
10. Prefontaine\\
11. The English Patient\\
12. Love and Basketball\\
13. The Fox and the Hound\\
14. The Cable Guy\\
15. Two Family House\\
16. Star Wars: Episode IV - A New Hope\\
17. Weekend at Bernie"s\\
18. Seven Days in May\\
19. 8 Seconds\\
20. Kurt \& Courtney\\ 
\midrule

{\color{blue}\textbf{Test User:}} \\\\
Learn from the above demonstration examples to solve the following test example.\\

\textbf{The User's Movie Profile}:\\
- Watched Movies: {\color{violet}[`0. Nikita (La Femme Nikita)', `1. Wallace \& Gromit: The Best of Aardman Animation', `2. Unforgiven', `3. Batman: Mask of the Phantasm', `4. The Fugitive', `5. Out of Sight', `6. Twelve Monkeys', `7. GoodFellas', `8. Fight Club', `9. Groundhog Day', `10. A Bug"s Life', `11. Tombstone', `12. Aladdin', `13. Beauty and the Beast', `14. Grosse Pointe Blank', `15. Election', `16. Leaving Las Vegas', `17. Total Recall', `18. A Few Good Men', `19. Pleasantville', `20. Jerry Maguire', `21. Pretty Woman', `22. Contact', `23. True Lies', `24. Waking Ned Devine', `25. Romeo Must Die', `26. Mission: Impossible 2', `27. Mission to Mars', `28. Killer, The (Die xue shuang xiong)', `29. Blade Runner', `30. The Princess Bride', `31. Brazil', `32. Henry V', `33. Amadeus', `34. The Right Stuff', `35. The Terminator', `36. Stand by Me', `37. Back to the Future', `38. This Is Spinal Tap', `39. Gandhi', `40. Star Trek: The Wrath of Khan', `41. Ghostbusters', `42. Mad Max 2 (a.k.a. The Road Warrior)', `43. A Fish Called Wanda', `44. Trading Places', `45. Chariots of Fire', `46. Time Bandits', `47. Who Framed Roger Rabbit?', `48. Indiana Jones and the Last Crusade', `49. Die Hard']}

\textbf{The User's Potential Matches}:\\
- Candidate Movies: {\color{violet}[`0. Sorority House Massacre', `1. Arguing the World', `2. Cop Land', `3. The Inkwell', `4. Lovers of the Arctic Circle, The (Los Amantes del Círculo Polar)', `5. Airplane!', `6. Pokémon the Movie 2000', `7. Johnny 100 Pesos', `8. Happy, Texas', `9. Arlington Road', `10. The Prophecy II', `11. Prom Night', `12. The Commitments', `13. Federal Hill', `14. Another Stakeout', `15. The Postman Always Rings Twice', `16. The Crow', `17. Shanghai Surprise', `18. Patriot Games', `19. Modulations']}

Based on the user's watched movies, please rank the candidate movies that align closely with John's preferences. \\
- You ONLY rank the given Candidate Movies.\\
- You DO NOT generate movies from Watched Movies.\\

Present your response in the format below: 1. [Top Recommendation (Candidate Movie)] \,\, 2. [2nd Recommendation (Candidate Movie)] \,\, 
...\,\,
20. [20th Recommendation (Candidate Movie)]\\
\textbf{Answer}:\\
1. The Commitments\\
2. Cop Land\\
3. {\fontfamily{lmtt}\fontseries{b}\selectfont \color{Maroon} Airplane!}\\
4. The Crow\\
5. Patriot Games\\
6. The Postman Always Rings Twice\\
7. Shanghai Surprise\\
8. Arlington Road\\
9. Happy, Texas\\
10. Another Stakeout\\
11. The Inkwell\\
12. Modulations\\
13. Johnny 100 Pesos\\
14. Sorority House Massacre\\
15. Pokémon the Movie 2000\\
16. Lovers of the Arctic Circle, The (Los Amantes del Círculo Polar)\\
17. The Prophecy II\\
18. Federal Hill\\
19. Prom Night\\
20. Arguing the World\\
\bottomrule
\end{longtable}
\end{center}
\end{spacing}
\rmfamily
\normalsize

\newpage
\begin{spacing}{1.0}
\begin{center}
\ttfamily
\footnotesize
\begin{longtable}{p{0.96\linewidth}}
    \caption[Example of LLMSRec-Nearest on the ML-1M dataset]{Example of the one-sho LLMSRec-Nearest on the ML-1M dataset.} \label{chap_5:case2} \\ \toprule
    {\color{blue}\textbf{Nearest Demonstration Example:}} \\\\
\textbf{The User's Movie Profile}:\\
- Watched Movies: {\color{violet}[`E.T. the Extra-Terrestrial', `Gladiator', `Raiders of the Lost Ark', `Brazil', `Aliens', `Full Metal Jacket', `The Right Stuff', `The Terminator', `Down by Law', `Blade Runner', `The Princess Bride', `Mystery Train', `Stand by Me', `Dangerous Liaisons', `Year of Living Dangerously', `Poltergeist', `Crimes and Misdemeanors', `Never Cry Wolf', `Mad Max 2 (a.k.a. The Road Warrior)', `Women on the Verge of a Nervous Breakdown', "Ferris Bueller's Day Off", `Who Framed Roger Rabbit?', `Koyaanisqatsi', `Ghostbusters', `A Fish Called Wanda', `The Gods Must Be Crazy', `Chariots of Fire', `The Hidden', `Diva', `Gandhi', `F/X', `Sex, Lies, and Videotape', `Reds', `Back to the Future', `Driving Miss Daisy', `Body Heat', `Indiana Jones and the Last Crusade', `When Harry Met Sally...', `Batman', `The Untouchables', `Predator', `Risky Business', `The Blues Brothers', `Star Trek IV: The Voyage Home', `U2: Rattle and Hum', `No Way Out', `The Color Purple', `Romancing the Stone', `The Color of Money', `Caddyshack']}\\

\textbf{The User's Potential Matches}:\\
- Candidate Movies: {\color{violet}[`Children of a Lesser God', `Kicking and Screaming', `Harlem', `The Good Mother', `Oscar and Lucinda (a.k.a. Oscar \& Lucinda)', `Meet Me in St. Louis', `Sinbad and the Eye of the Tiger', `Gremlins', "Freddy's Dead: The Final Nightmare", `One Magic Christmas', `The Scarlet Letter', `Alaska', `Red Firecracker, Green Firecracker', `Bulletproof', `The Crimson Pirate', `Clockers', `The Stepford Wives', `Twelve Monkeys', `Bitter Moon', `Friday the 13th: The Final Chapter']}\\
Based on the user's watched movies, please rank the candidate movies that align closely with the user's preferences. \\
- You ONLY rank the given Candidate Movies.\\
- You DO NOT generate movies from Watched Movies.\\

Present your response in the format below:\\ 
1. [Top Recommendation (Candidate Movie)] \\ 
2. [2nd Recommendation (Candidate Movie)]\\ 
...\\
20. [20th Recommendation (Candidate Movie)]\\
\textbf{Answer}:\\
1. Children of a Lesser God\\
2. One Magic Christmas\\
3. Alaska\\
4. Gremlins\\
5. The Good Mother\\
6. Bitter Moon\\
7. Bulletproof\\
8. Freddy's Dead: The Final Nightmare\\
9. Red Firecracker, Green Firecracker\\
10. The Scarlet Letter\\
11. Sinbad and the Eye of the Tiger\\
12. Kicking and Screaming\\
13. Meet Me in St. Louis\\
14. Oscar and Lucinda (a.k.a. Oscar \& Lucinda)\\
15. Twelve Monkeys\\
16. Clockers\\
17. Harlem\\
18. The Stepford Wives\\
19. The Crimson Pirate\\
20. Friday the 13th: The Final Chapter\\
\midrule

{\color{blue}\textbf{Test User:}} \\\\
Learn from the above demonstration examples to solve the following test example.\\

\textbf{The User's Movie Profile}:\\
- Watched Movies: {\color{violet}[`0. Nikita (La Femme Nikita)', `1. Wallace \& Gromit: The Best of Aardman Animation', `2. Unforgiven', `3. Batman: Mask of the Phantasm', `4. The Fugitive', `5. Out of Sight', `6. Twelve Monkeys', `7. GoodFellas', `8. Fight Club', `9. Groundhog Day', `10. A Bug"s Life', `11. Tombstone', `12. Aladdin', `13. Beauty and the Beast', `14. Grosse Pointe Blank', `15. Election', `16. Leaving Las Vegas', `17. Total Recall', `18. A Few Good Men', `19. Pleasantville', `20. Jerry Maguire', `21. Pretty Woman', `22. Contact', `23. True Lies', `24. Waking Ned Devine', `25. Romeo Must Die', `26. Mission: Impossible 2', `27. Mission to Mars', `28. Killer, The (Die xue shuang xiong)', `29. Blade Runner', `30. The Princess Bride', `31. Brazil', `32. Henry V', `33. Amadeus', `34. The Right Stuff', `35. The Terminator', `36. Stand by Me', `37. Back to the Future', `38. This Is Spinal Tap', `39. Gandhi', `40. Star Trek: The Wrath of Khan', `41. Ghostbusters', `42. Mad Max 2 (a.k.a. The Road Warrior)', `43. A Fish Called Wanda', `44. Trading Places', `45. Chariots of Fire', `46. Time Bandits', `47. Who Framed Roger Rabbit?', `48. Indiana Jones and the Last Crusade', `49. Die Hard']}

\textbf{The User's Potential Matches}:\\
- Candidate Movies: {\color{violet}[`0. Sorority House Massacre', `1. Arguing the World', `2. Cop Land', `3. The Inkwell', `4. Lovers of the Arctic Circle, The (Los Amantes del Círculo Polar)', `5. Airplane!', `6. Pokémon the Movie 2000', `7. Johnny 100 Pesos', `8. Happy, Texas', `9. Arlington Road', `10. The Prophecy II', `11. Prom Night', `12. The Commitments', `13. Federal Hill', `14. Another Stakeout', `15. The Postman Always Rings Twice', `16. The Crow', `17. Shanghai Surprise', `18. Patriot Games', `19. Modulations']}

Based on the user's watched movies, please rank the candidate movies that align closely with John's preferences. \\
- You ONLY rank the given Candidate Movies.\\
- You DO NOT generate movies from Watched Movies.\\

Present your response in the format below: 1. [Top Recommendation (Candidate Movie)] \,\, 2. [2nd Recommendation (Candidate Movie)] \,\, 
...\,\,
20. [20th Recommendation (Candidate Movie)]\\
\textbf{Answer}:\\
1. Arlington Road\\
2. Cop Land\\
3. The Crow\\
4. Patriot Games\\
5. The Postman Always Rings Twice\\
6. The Commitments\\
7. {\fontfamily{lmtt}\fontseries{b}\selectfont \color{Maroon}Airplane!}\\
8. Another Stakeout\\
9. Lovers of the Arctic Circle, The (Los Amantes del Círculo Polar)\\
10. Shanghai Surprise\\
11. Happy, Texas\\
12. Modulations\\
13. The Inkwell\\
14. Johnny 100 Pesos\\
15. Sorority House Massacre\\
16. Arguing the World\\
17. Prom Night\\
18. Federal Hill\\
19. Pokémon the Movie 2000\\
20. The Prophecy II\\ 
\bottomrule
\end{longtable}
\end{center}
\end{spacing}
\rmfamily
\normalsize

\newpage
\onecolumn
\begin{spacing}{1.0}
\begin{center}
\ttfamily
\footnotesize
\begin{longtable}{p{0.96\linewidth}}
    \caption[Example of LLMSRec-Fixed on the ML-1M dataset]{Example of the one-sho LLMSRec-Fixed on the ML-1M dataset.} \label{chap_5:case3} \\ \toprule
    {\color{blue}\textbf{Fixed Demonstration Example:}} \\\\
\textbf{The User's Movie Profile}:\\
- Watched Movies: {\color{violet}[`Total Recall', `Aliens', `Star Wars: Episode VI - Return of the Jedi', `E.T. the Extra-Terrestrial', `Forbidden Planet', `Brazil', `Star Trek: First Contact', `Star Trek: The Wrath of Khan', `Sneakers', `Galaxy Quest', `Contact', `Village of the Damned', `Being John Malkovich', `Waiting for Guffman', `Clerks', `American Beauty', `Toy Story 2', `Shakespeare in Love', `Toy Story', `Flirting With Disaster', `Smoke Signals', `Pulp Fiction', `Erin Brockovich', `Chicken Run', `Shanghai Noon', `Gladiator', `The Wizard of Oz', `The Producers', "Singin' in the Rain", `The Sound of Music', `Snow White and the Seven Dwarfs', `Fantasia', `Sleeping Beauty', `West Side Story', `Cinderella', `The Little Mermaid', `Holiday Inn', `James and the Giant Peach', `Dumbo', `The Lion King', `The Nightmare Before Christmas', `The Rocky Horror Picture Show', `Oliver!', `Pocahontas', `Everyone Says I Love You', `South Pacific', `Victor/Victoria', `Tank Girl', `His Girl Friday', `The Maltese Falcon']}\\

\textbf{The User's Potential Matches}:\\
- Candidate Movies: {\color{violet}[`The Shop Around the Corner',`Kicking and Screaming',`Harlem',`The Good Mother',`Oscar and Lucinda (a.k.a. Oscar \& Lucinda)',`Roman Holiday',`Sinbad and the Eye of the Tiger',`Gremlins', "Freddy's Dead: The Final Nightmare",`One Magic Christmas',`The Scarlet Letter',`Alaska',`Red Firecracker, Green Firecracker',`Bulletproof',`The Crimson Pirate',`Clockers',`The Stepford Wives',`The Crow',`Bitter Moon',`Friday the 13th: The Final Chapter']}\\
Based on the user's watched movies, please rank the candidate movies that align closely with the user's preferences. \\
- You ONLY rank the given Candidate Movies.\\
- You DO NOT generate movies from Watched Movies.\\

Present your response in the format below:\\ 
1. [Top Recommendation (Candidate Movie)] \\ 
2. [2nd Recommendation (Candidate Movie)]\\ 
...\\
20. [20th Recommendation (Candidate Movie)]\\
\textbf{Answer}:\\
1. The Shop Around the Corner\\
2. One Magic Christmas\\
3. Alaska\\
4. Gremlins\\
5. The Good Mother\\
6. Bitter Moon\\
7. Bulletproof\\
8. Freddy's Dead: The Final Nightmare\\
9. Red Firecracker, Green Firecracker\\
10. The Scarlet Letter\\
11. Sinbad and the Eye of the Tiger\\
12. Kicking and Screaming\\
13. Roman Holiday\\
14. Oscar and Lucinda (a.k.a. Oscar \& Lucinda)\\
15. The Crow\\
16. Clockers\\
17. Harlem\\
18. The Stepford Wives\\
19. The Crimson Pirate\\
20. Friday the 13th: The Final Chapter\\
\midrule

{\color{blue}\textbf{Test User:}} \\\\
Learn from the above demonstration examples to solve the following test example.\\

\textbf{The User's Movie Profile}:\\
- Watched Movies: {\color{violet}[`0. Nikita (La Femme Nikita)', `1. Wallace \& Gromit: The Best of Aardman Animation', `2. Unforgiven', `3. Batman: Mask of the Phantasm', `4. The Fugitive', `5. Out of Sight', `6. Twelve Monkeys', `7. GoodFellas', `8. Fight Club', `9. Groundhog Day', `10. A Bug"s Life', `11. Tombstone', `12. Aladdin', `13. Beauty and the Beast', `14. Grosse Pointe Blank', `15. Election', `16. Leaving Las Vegas', `17. Total Recall', `18. A Few Good Men', `19. Pleasantville', `20. Jerry Maguire', `21. Pretty Woman', `22. Contact', `23. True Lies', `24. Waking Ned Devine', `25. Romeo Must Die', `26. Mission: Impossible 2', `27. Mission to Mars', `28. Killer, The (Die xue shuang xiong)', `29. Blade Runner', `30. The Princess Bride', `31. Brazil', `32. Henry V', `33. Amadeus', `34. The Right Stuff', `35. The Terminator', `36. Stand by Me', `37. Back to the Future', `38. This Is Spinal Tap', `39. Gandhi', `40. Star Trek: The Wrath of Khan', `41. Ghostbusters', `42. Mad Max 2 (a.k.a. The Road Warrior)', `43. A Fish Called Wanda', `44. Trading Places', `45. Chariots of Fire', `46. Time Bandits', `47. Who Framed Roger Rabbit?', `48. Indiana Jones and the Last Crusade', `49. Die Hard']}

\textbf{The User's Potential Matches}:\\
- Candidate Movies: {\color{violet}[`0. Sorority House Massacre', `1. Arguing the World', `2. Cop Land', `3. The Inkwell', `4. Lovers of the Arctic Circle, The (Los Amantes del Círculo Polar)', `5. Airplane!', `6. Pokémon the Movie 2000', `7. Johnny 100 Pesos', `8. Happy, Texas', `9. Arlington Road', `10. The Prophecy II', `11. Prom Night', `12. The Commitments', `13. Federal Hill', `14. Another Stakeout', `15. The Postman Always Rings Twice', `16. The Crow', `17. Shanghai Surprise', `18. Patriot Games', `19. Modulations']}

Based on the user's watched movies, please rank the candidate movies that align closely with John's preferences. \\
- You ONLY rank the given Candidate Movies.\\
- You DO NOT generate movies from Watched Movies.\\

Present your response in the format below: 1. [Top Recommendation (Candidate Movie)] \,\, 2. [2nd Recommendation (Candidate Movie)] \,\, 
...\,\,
20. [20th Recommendation (Candidate Movie)]\\
\textbf{Answer}:\\
1. The Crow\\
2. Shanghai Surprise\\
3. The Commitments\\
4. Arlington Road\\
5. The Postman Always Rings Twice\\
6. Lovers of the Arctic Circle, The (Los Amantes del Círculo Polar)\\
7. Cop Land\\
8. Happy, Texas\\
9. Another Stakeout\\
10. {\fontfamily{lmtt}\fontseries{b}\selectfont \color{Maroon}Airplane!}\\
11. Federal Hill\\
12. Johnny 100 Pesos\\
13. The Prophecy II\\
14. Modulations\\
15. Patriot Games\\
16. Pokémon the Movie 2000\\
17. Sorority House Massacre\\
18. Prom Night\\
19. Arguing the World\\
20. The Inkwell\\
\bottomrule
\end{longtable}
\end{center}
\end{spacing}
\rmfamily
\normalsize

\newpage
\onecolumn
\begin{spacing}{1.0}
\begin{center}
\ttfamily
\footnotesize
\begin{longtable}{p{0.96\linewidth}}
    \caption[Example of zero-shot LLMSRec on the ML-1M dataset]{Example of the zero-shot LLMSRec on the ML-1M dataset.} \label{chap_5:case4} \\ \toprule

{\color{blue}\textbf{Test User:}} \\\\
Learn from the above demonstration examples to solve the following test example.\\

\textbf{The User's Movie Profile}:\\
- Watched Movies: {\color{violet}[`0. Nikita (La Femme Nikita)', `1. Wallace \& Gromit: The Best of Aardman Animation', `2. Unforgiven', `3. Batman: Mask of the Phantasm', `4. The Fugitive', `5. Out of Sight', `6. Twelve Monkeys', `7. GoodFellas', `8. Fight Club', `9. Groundhog Day', `10. A Bug"s Life', `11. Tombstone', `12. Aladdin', `13. Beauty and the Beast', `14. Grosse Pointe Blank', `15. Election', `16. Leaving Las Vegas', `17. Total Recall', `18. A Few Good Men', `19. Pleasantville', `20. Jerry Maguire', `21. Pretty Woman', `22. Contact', `23. True Lies', `24. Waking Ned Devine', `25. Romeo Must Die', `26. Mission: Impossible 2', `27. Mission to Mars', `28. Killer, The (Die xue shuang xiong)', `29. Blade Runner', `30. The Princess Bride', `31. Brazil', `32. Henry V', `33. Amadeus', `34. The Right Stuff', `35. The Terminator', `36. Stand by Me', `37. Back to the Future', `38. This Is Spinal Tap', `39. Gandhi', `40. Star Trek: The Wrath of Khan', `41. Ghostbusters', `42. Mad Max 2 (a.k.a. The Road Warrior)', `43. A Fish Called Wanda', `44. Trading Places', `45. Chariots of Fire', `46. Time Bandits', `47. Who Framed Roger Rabbit?', `48. Indiana Jones and the Last Crusade', `49. Die Hard']}

\textbf{The User's Potential Matches}:\\
- Candidate Movies: {\color{violet}[`0. Sorority House Massacre', `1. Arguing the World', `2. Cop Land', `3. The Inkwell', `4. Lovers of the Arctic Circle, The (Los Amantes del Círculo Polar)', `5. Airplane!', `6. Pokémon the Movie 2000', `7. Johnny 100 Pesos', `8. Happy, Texas', `9. Arlington Road', `10. The Prophecy II', `11. Prom Night', `12. The Commitments', `13. Federal Hill', `14. Another Stakeout', `15. The Postman Always Rings Twice', `16. The Crow', `17. Shanghai Surprise', `18. Patriot Games', `19. Modulations']}

Based on the user's watched movies, please rank the candidate movies that align closely with John's preferences. \\
- You ONLY rank the given Candidate Movies.\\
- You DO NOT generate movies from Watched Movies.\\

Present your response in the format below: 1. [Top Recommendation (Candidate Movie)] \,\, 2. [2nd Recommendation (Candidate Movie)] \,\, 
...\,\,
20. [20th Recommendation (Candidate Movie)]\\
\textbf{Answer}:\\
1. The Commitments\\
2. Patriot Games\\
3. Arlington Road\\
4. Cop Land\\
5. The Crow\\
6. The Postman Always Rings Twice\\
7. Shanghai Surprise\\
8. Another Stakeout\\
9. Modulations\\
10. Pokémon the Movie 2000\\
11. The Inkwell\\
12. Happy, Texas\\
13. The Prophecy II\\
14. Johnny 100 Pesos\\
15. Lovers of the Arctic Circle, The (Los Amantes del Círculo Polar)\\
16. Arguing the World\\
17. Federal Hill\\
18. Prom Night\\
19. Sorority House Massacre\\
20. {\fontfamily{lmtt}\fontseries{b}\selectfont \color{Maroon}Airplane!}\\
\bottomrule
\end{longtable}
\end{center}
\end{spacing}
\rmfamily
\normalsize


\label{sec:appendix}

\end{document}